\pdfoutput=1

\documentclass[11pt,twoside,a4paper,cmspaper,final,collab]{cms-tdr}
\def\svnVersion{425872P}\def\cmsCernNoTag{CERN-EP-2016-321}\def\cmsCernDate{\today}\def\cmsMessage{Published in the Journal of High Energy Physics as \href{http://dx.doi.org/10.1007/JHEP09(2017)051}{\doi{10.1007/JHEP09(2017)051.}}}

\begin{document}\cmsNoteHeader{TOP-16-006}

\hyphenation{had-ron-i-za-tion}
\hyphenation{cal-or-i-me-ter}
\hyphenation{de-vices}
\RCS$Revision: 421789 $
\RCS$HeadURL: svn+ssh://svn.cern.ch/reps/tdr2/papers/TOP-16-006/trunk/TOP-16-006.tex $
\RCS$Id: TOP-16-006.tex 421789 2017-08-18 19:12:07Z psilva $

\cmsNoteHeader{TOP-16-006}
\title{Measurement of the \ttbar production cross section
  using events with one lepton and at least one jet in pp collisions at $\sqrt{s}=13$\TeV}

\date{\today}

\abstract{
  A measurement of the $\ttbar$ production cross section at
  $\sqrt{s}=13$\TeV is presented using proton-proton
  collisions, corresponding to an integrated luminosity of 2.2\fbinv,
  collected with the CMS detector at the LHC.
  Final states with one isolated charged lepton (electron or muon) and at
  least one jet are selected and categorized according to the accompanying jet
  multiplicity. From a likelihood fit to the invariant mass distribution of
  the isolated lepton and a jet identified as coming from the
  hadronization of a bottom quark, the cross section is
  measured to be
  $\sigma_{\ttbar}= 888 \pm 2\stat ^{+26}_{-28}\syst\pm 20\lum\unit{pb}$,
  in agreement with the standard model prediction.
  Using the expected dependence of the cross section on the pole mass
  of the top quark ($m_{\PQt}$), the value of $m_{\PQt}$ is found
  to be $170.6\pm2.7$\GeV.
}

\hypersetup{%
pdfauthor={CMS Collaboration},%
pdftitle={Measurement of the ttbar production cross section using
  events with one lepton and at least one jet in pp collisions at sqrt(s)=13 TeV},%
pdfsubject={CMS},%
pdfkeywords={CMS, top, cross section, pp collisions}}

\maketitle

\section{Introduction}
\label{sec:intro}

The rate at which top quark-antiquark (\ttbar) pairs are produced in
proton-proton (pp) collisions at LHC has been measured at
 center-of-mass energies of 7~\cite{ATLAS:2012aa,Aad:2011yb,Aad:2010ey,Aad:2012mza,Khachatryan:2010ez,Chatrchyan:2011nb,Chatrchyan:2011ew,Chatrchyan:2011vp,Chatrchyan:2012vs,Chatrchyan:2012bra,Chatrchyan:2012cz,Chatrchyan:2013kff,Chatrchyan:2012ria,Chatrchyan:2013ual}, 8~\cite{Aad:2014kva,Aad:2015pga,Chatrchyan:2013faa,Khachatryan:2014loa,Khachatryan:2015fwh,Khachatryan:2016yzq,Aaij:2015mwa,Aaij:2016vsy,Khachatryan:2016mqs}, and
13\TeV~\cite{Khachatryan:2015uqb,Aaboud:2016pbd,Khachatryan:2016kzg}.
The latter has been determined experimentally with a 4.4\% uncertainty.
In addition, several analyses
have explored the expected dependence of the
\ttbar production
cross section ($\sigma_{\ttbar}$)
on the mass of the top quark ($m_\PQt$)
to extract the latter.
Recent examples of this can be found in
Ref.~\cite{Khachatryan:2016mqs}, where
$m_\PQt$ is determined with a total uncertainty of $\approx$1\%.
Alternatively, the strong coupling strength ($\alpha_\mathrm{S}$) can be extracted from the $\ttbar$ cross section, assuming
$m_\PQt$ is known~\cite{Chatrchyan:2013haa}.
Knowledge of the parton distribution function (PDF) of the proton can be
improved as well from a precise measurement of $\sigma_{\ttbar}$~\cite{Ball:2014uwa,Harland-Lang:2014zoa}.
In addition, the production of final states via processes beyond the standard model that mimic the ones produced
by \ttbar decay can be revealed by a precise measurement of $\sigma_{\ttbar}$~\cite{Czakon:2014fka}.
The above-mentioned interpretations of the measured $\sigma_{\ttbar}$
provide a few examples, among others existing in the literature,
that can benefit from such precision comparisons.

In this paper, a measurement of $\sigma_{\ttbar}$
using final states with an isolated charged lepton $\ell$ (electron or muon)
and at least one jet is presented.
This selection is chosen in order to minimize the uncertainty
in the extrapolation of the cross section to the fully inclusive phase
space,
and is expected to keep the impact of the dependence of the
acceptance
on the theoretical uncertainties in the PDFs
and quantum chromodynamics (QCD) scale choice to a minimum.
The selected events are split into categories
according to the total number of jets in the event and the number of jets identified
as coming from the hadronization of a \cPqb{} quark.
Each category uses observables that can discriminate the
main backgrounds (multijet and \PW+jets production)
from the \ttbar signal. A combined fit to the distributions in data of
these observables is used to minimize the main
systematic uncertainties, while measuring $\sigma_{\ttbar}$ and $m_\PQt$.

The paper is organized as follows: Section~\ref{sec:experimentsetup}
details the experimental setup, including the CMS detector, the data and simulation used in the analysis,
the event selection, and the background
estimations,
Section~\ref{sec:fit} describes the observables used in the analysis
and the associated systematic uncertainties,
while Section~\ref{sec:results} discusses the fit procedure and results.
A summary is given in Section~\ref{sec:summary}.

\section{Experimental setup}
\label{sec:experimentsetup}

\subsection{The CMS detector}
\label{subsec:cmsdet}

The central feature of the CMS apparatus is a superconducting solenoid
of 6\unit{m} internal diameter, providing a magnetic field of 3.8\unit{T}.
Within the solenoid volume are a silicon pixel and strip tracker, a
lead tungstate crystal electromagnetic calorimeter,
and a brass and scintillator hadron calorimeter,
each composed of a barrel and two endcap sections.
Forward calorimeters extend the pseudorapidity ($\eta$) coverage provided by
the barrel and endcap detectors.
Muons are detected in gas-ionization chambers embedded in the steel flux-return yoke outside the solenoid.
A more detailed description of the CMS detector, together with a
definition of the coordinate system
used and the relevant kinematic variables, can be found in Ref.~\cite{Chatrchyan:2008zzk}.

\subsection{Data and simulation}
\label{subsec:datasim}

The analysis is based on pp collision data collected by the
CMS experiment at the CERN LHC at $\sqrt{s} = 13$\TeV in 2015, corresponding to an
integrated luminosity of $2.21 \pm0.05\fbinv$~\cite{CMS-PAS-LUM-15-001}.

The analysis is complemented using simulated event samples that are used
to estimate the main backgrounds and the signal distributions.
The \ttbar signal is modeled with the \POWHEG~v2~\cite{powheg1,powheg2,powheg3,Frixione:2007nw} generator,
matched to \PYTHIA~v8.205~\cite{Sjostrand:2006za,Sjostrand:2014zea} for shower evolution and hadronization.
The NNPDF3.0 next-to-leading-order (NLO) PDFs~\cite{Demartin:2010er} and the
{CUETP8M1}~\cite{Khachatryan:2015pea,Skands:2014pea} underlying-event tune are used in the simulation.
To evaluate the systematic uncertainties associated with the QCD
renormalization ($\mu_\mathrm{R}$) and factorization ($\mu_\mathrm{F}$) scales
at the matrix-element level, we make use
of a weighting scheme implemented in \POWHEG~v2
to vary the scales by a factor of 2 or 1/2
relative to its nominal value $\mu_\mathrm{R}=\mu_\mathrm{F}=m_\mathrm{T}$,
where $m_\mathrm{T}=\sqrt{\smash[b]{m_\PQt^2+p_{\mathrm{T},\PQt}^2}}$ is the transverse mass of the
top quark, with $p_{\mathrm{T},\PQt}$ being the top quark transverse momentum.

Furthermore, additional simulations in which the QCD renormalization and factorization scales
at the parton shower level are changed by a factor of 2 or 1/2
relative to their nominal value are used.
In the {CUETP8M1} tune, the nominal QCD scale choice at the parton shower level is
determined by $\alpha_\mathrm{S}^\mathrm{ISR}=0.1365$,
the value of the strong coupling strength at $m_\cPZ$ used for the initial-state shower.
A different matrix-element generator is also used, for comparison:
\textsc{MG5}\_a\MCATNLO~v5\_2.2.2~\cite{Alwall:2014hca} with
\textsc{Madspin}~\cite{madspin}, and is matched to either \PYTHIA~8 or
\HERWIGpp~v2.7.1~\cite{herwigpp}.

{\tolerance=1200
In this analysis, we measure the \ttbar cross section in a fiducial region of
the phase space using as reference
the theoretical cross section for  $m_\PQt=172.5\GeV$,
computed at next-to-next-to-leading order (NNLO) with
next-to-next-to-leading-log (NNLL) soft-gluon resummations,
 $\sigma_{\ttbar} = 832~^{+20}_{-29}\,(\text{scale})\pm
 35\,(\mathrm{PDF}+\alpha_\mathrm{S})\unit{pb}$,
from~\textsc{top++}~v2.0~\cite{top++}.
Single top quark processes are simulated with \POWHEG~v1~\cite{Alioli:2009je,Re:2010bp}
and normalized to the approximate NNLO prediction~\cite{Kidonakis:2013zqa}.
The \PW+jets process is simulated at NLO with \textsc{MG5}\_a\MCATNLO.
To reach higher statistical accuracy, a larger Born-level
\MADGRAPH~v5.1.3.30~\cite{Alwall:2014hca} simulated sample,
including up to four extra partons in the matrix-element calculations, is used for the
derivation of the \PW+jets background shape.
The Drell--Yan (DY) contribution is simulated with \MADGRAPH.
Both \PW+jets and DY cross sections are normalized to their NNLO
predictions, computed using {\FEWZ} (v3.1.b2)~\cite{fewz}.
Diboson production (WW, ZZ, WZ) is simulated either with \PYTHIA8 (\cPZ\cPZ, \PW\cPZ) or
\POWHEG~v1 \cite{Melia:2011tj} (\PW\PW).
Each diboson process is normalized to the NLO prediction
for the cross section, computed with \MCFM
(v7.0)~\cite{mcfm,Campbell:2011bn}.
The associated production of \PW{} or \cPZ{} boson with \ttbar
(\ttbar+V) is simulated at NLO with \textsc{MG5}\_a\MCATNLO.
\par}

All simulated events include an emulation of the response of the CMS detector
using \GEANTfour{} v9.4p03~\cite{GEANT4,Agostinelli:2002hh}.
The effect due to multiple pp collisions
in the same and neighboring beam crossings (pileup)
is measured and added to the simulated \ttbar interactions
according to the pileup multiplicity
observed in the data.

\subsection{Event selection}
\label{subsec:evsel}

The data are recorded using single-lepton triggers
with a minimum transverse momentum (\pt) of 22\GeV and 20\GeV for electrons and muons, respectively.
Identification and isolation criteria are applied at the trigger level, and
the efficiency of these requirements is measured in a control data
sample that is dominated by $\cPZ\to\ell\ell$ decays.
The results obtained from the control data sample are compared
with the simulated predictions using
a tag-and-probe method~\cite{Khachatryan:2010xn}, and
data-to-simulation scale correction factors are derived
as function of the \pt and $\eta$ of the lepton.
The scale factors are observed to be $\leq$5\%.

The events are reconstructed offline using a particle-flow (PF) algorithm that optimally combines the information from subdetectors
to reconstruct and identify all individual particles in the event~\cite{CMS-PRF-14-001}.
In addition, reconstruction, identification, and calibration algorithms are employed for electrons and muons, as described in Refs.~\cite{Khachatryan:2015hwa,Chatrchyan:2013sba}.
The lepton candidates are required to have $\pt>30\GeV$ and
$\abs{\eta}<2.1$.
Identification and isolation requirements are imposed to reject
misidentified muons from punchthrough hadrons, photon conversion, and other objects misreconstructed as
lepton candidates.
These criteria are tighter than the ones imposed at trigger level.
The tag-and-probe method measures the
efficiency of these requirements, yielding typical efficiencies of
70\% and 92\% for electrons and muons, respectively.
Nonprompt leptons that come from the decays of
long-lived hadrons are rejected by requiring that the significance of the three-dimensional (3D)
impact parameter of the lepton track, relative to the
primary event vertex, is less than four standard deviations.
This requirement effectively reduces the contamination
from multijet events, while keeping a high efficiency for the signal. The
expected efficiency of this requirement is cross-checked using
$\cPZ\to\ell\ell$ candidate events.
The primary event vertex used as reference is required to be reconstructed from at least four tracks,
and have a longitudinal distance of less than 24 cm from the center of the detector. Among all the pp collision
vertices in the event, the one with the largest scalar sum of associated particle transverse momenta is selected
as the primary vertex.
The event is rejected if an additional electron or
muon is found within  $\abs{\eta}\leq2.5$, passing looser identification and isolation criteria,
and  with $\pt>15$ or 10\GeV, respectively.

Jets are reconstructed using all PF candidates as
inputs to the anti-\kt algorithm with a
distance parameter of 0.4, utilizing the \FASTJET 3.1
software package~\cite{Cacciari:2008gp,Cacciari:2011ma}.
The jet momentum is defined as the vectorial sum of all particle momenta inside the jet cone, and is
found from the simulation to be within 5--10\% of the generated jet momentum at particle level over the whole \pt range and detector acceptance.
Since pileup collisions result in unwanted calorimetric energy depositions and extra tracks,
part of this contribution is reduced by
performing a charged-hadron subtraction that
removes tracks identified as originating from pileup vertices~\cite{Khachatryan:2016kdb}.
In addition, an offset correction is applied to remove the additional energy
included in the jets that come from
pileup~\cite{Cacciari:2008gn,Cacciari:2007fd}.
The energy scale corrections,
derived from simulation, are cross-checked with in situ measurements of
the energy balance in dijet and photon+jet events~\cite{Khachatryan:2016kdb}.

We require at least one jet with $\pt>30\GeV$ and  $\abs{\eta}\leq
2.5$ in the accepted events. The jets are required to not overlap with the isolated
lepton within a cone of angular radius $\Delta
R=\sqrt{\smash[b]{(\Delta\eta)^2+(\Delta\phi)^2}}< 0.4$,
where $\Delta\eta$ and $\Delta\phi$, represent the difference in
pseudorapidity and azimuthal angle (in radians), between the directions of each jet and the lepton.
Jets coming from the fragmentation and hadronization of \cPqb{} quarks
(b jets) are identified by a combined secondary vertex (CSV) algorithm~\cite{Chatrchyan:2012jua}.
A b jet is identified with a CSV threshold efficiency $>$65\% and a
misidentification rate $\approx$1\%.
This \cPqb{} tagging efficiency is measured using a \bbbar
enriched data sample from a method similar to that described in Ref.~\cite{Chatrchyan:2012jua}.

In the analysis, events with one, two, three, or four or more jets  are
considered as separate event categories. We expect the low-multiplicity
categories to be dominated by \PW+jets processes, and the high jet multiplicities by \ttbar events.
An additional separation of the signal is achieved by counting the
number of \cPqb-tagged jets in each category, since  two \cPqb{} jets
in the event are expected, given that each top quark decays to a \PW\cPqb{} pair.
Therefore, we further subdivide the four jet-multiplicity
categories according to the number of reconstructed \cPqb-tagged jets,
considering events with none, one, or at least two \cPqb-tagged jets, for a total of 11
orthogonal categories.
Since the collision particles are protons, an asymmetric
production of \PW{} bosons, with more $\PWp$ produced than
$\PWm$, is expected~\cite{Campbell:1999ah}. Given the charge-symmetric decays of the
$\PW$ bosons in \ttbar decays, \ttbar final states are expected to have the same number
of $\PWp$ and $\PWm$ bosons. We use this property to further
categorize the events according to the lepton charge ($+-$) and flavor (electron or muon).
Hence, our analysis makes use of a total of $2{\times}2{\times}11 = 44$ categories.

All backgrounds are estimated using simulation except for
that from multijet events, which is difficult to model correctly from simulation
in the \ttbar phase-space region. The contribution from the multijet background is
estimated using an independent data control sample
where the prompt-lepton candidate passes the loose
trigger-isolation requirements,
but fails the tighter isolation required offline.
The expected residual contamination from background
processes other than multijets is estimated from simulation and subtracted from the
control sample. The resulting
distributions are used to model the multijet
background contribution. The initial multijet normalization is
obtained from
events containing one isolated
lepton and having the measured
absolute value of the imbalance in the \pt
of all PF candidates in the event less than $20\GeV$.
The contributions from backgrounds other than multijets are subtracted
in the referred to isolated-lepton region, and the ratio of
events observed in data in this region with respect to the
number of events found in the nonisolated-lepton control region is assigned as the renormalization scale factor.
Given the tight requirements on leptons, we expect \bbbar+jets events to dominate the
multijet contamination.
An isolated, prompt lepton coming from such a process
is likely to arise from the decay of a bottom hadron.
We can therefore expect a jet in the event to be \cPqb-tagged.
This motivates the initial normalization for the multijet
process through the  one-\cPqb-tagged-jet category.
However, for events with at least three jets, the \ttbar contribution is expected
to be nonnegligible, so the multijet process is estimated
from events without any \cPqb-tagged jets.

\begin{figure}[htbp]
\centering
\includegraphics[width=0.99\textwidth]{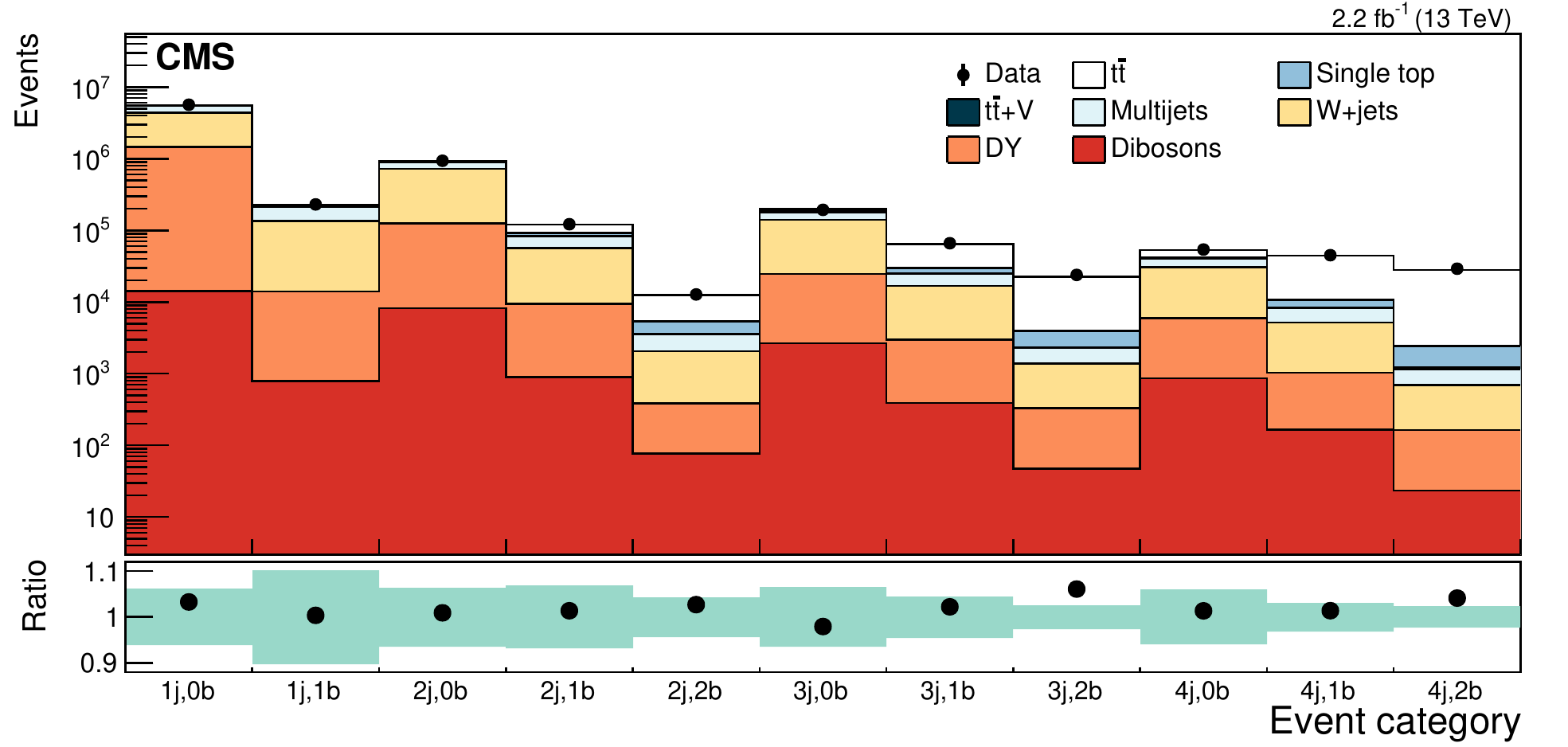}
\caption{Event yields from data and the expected
\ttbar signal and backgrounds for each of the 11 independent categories. Distributions are combined for the two lepton charges and flavors.
The bins represent the measured number of jets ({j}) and \cPqb-tagged jets
({b}), with the 4j and 2b categories being inclusive.
The bottom panel shows the ratio between the data and
the expectations. The relative uncertainty owing to the
statistical uncertainty in the simulation,
the uncertainty in the normalization of the
multijet contribution,
and  the systematic uncertainty in the total integrated
luminosity is represented as a shaded band.}
\label{fig:njetsnbtags}
\end{figure}

Figure~\ref{fig:njetsnbtags} compares the numbers of selected events
in data with the signal and expected backgrounds from simulation in
each category. For simplicity, the contributions from
the electron and muon final states, as well as from the two
lepton charges, are summed. Within the uncertainties, we observe agreement
between the data and the expectations.
Although not shown explicitly, agreement is also found separately for
each lepton flavor and charge.

\section{Observables and related uncertainties}
\label{sec:fit}

For each event category,
we select a variable that discriminates the signal from
the backgrounds.
Categories without \cPqb-tagged jets are likely to be dominated by
backgrounds and thus are counted without analyzing any distribution.
For events with \cPqb-tagged jets, we exploit the distinct kinematic character of
$\PQt\to\PW\cPqb$ decays,
and use the following mass variables:
(i)  for events with only one \cPqb-tagged jet, we use the invariant
mass of the system formed by the lepton and the \cPqb-tagged jet
($M(\ell,\cPqb)$); and (ii) for events with at least two \cPqb-tagged
jets, the invariant masses of all the lepton and \cPqb{}-tagged jet combinations in
the event are calculated, and the minimum mass $(\min M(\ell,\cPqb))$ is
chosen as a discriminant.
The $M(\ell,\cPqb)$-related variables are expected to be
sensitive to \ttbar production, as well as to $m_\PQt$, defined by the
endpoint in the invariant mass spectrum expected at leading order (LO).
The endpoint is determined by the values of the top quark and \PW{}
boson masses~\cite{Chatrchyan:2013boa}.

\begin{figure}[htbp]
\centering
\includegraphics[width=0.35\textwidth]{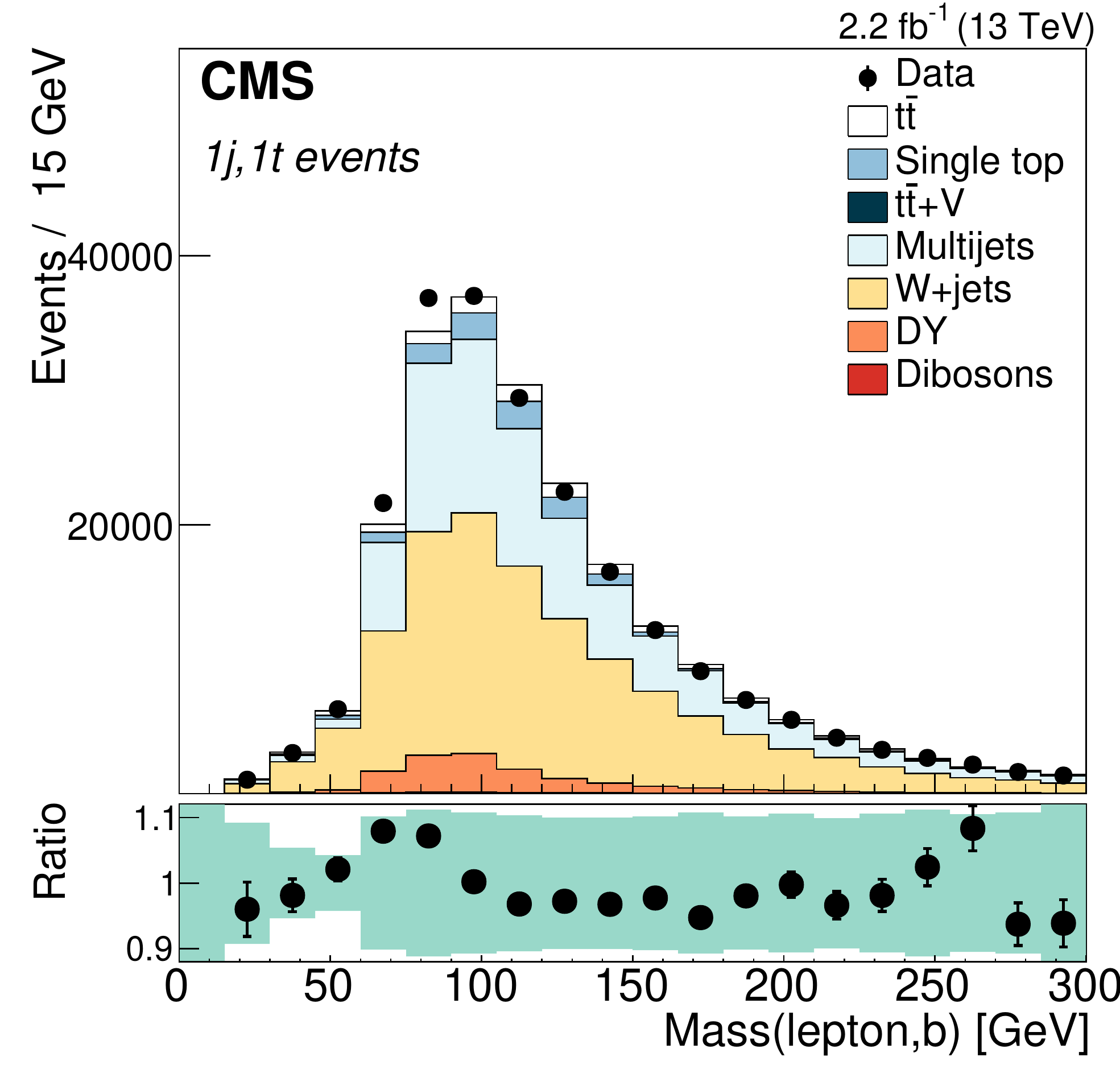}\hspace*{0.35\textwidth}~\\
\includegraphics[width=0.35\textwidth]{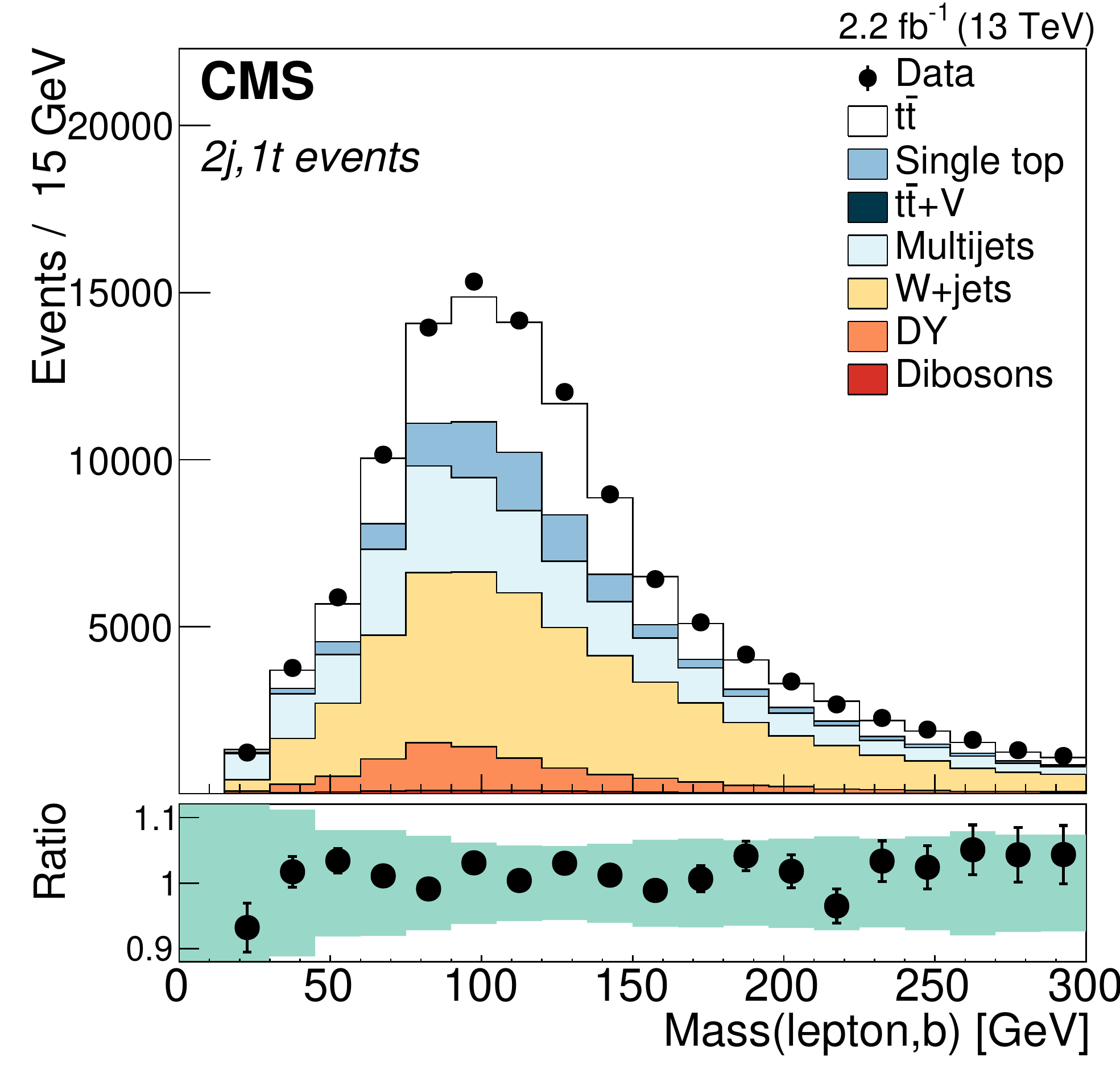}
\includegraphics[width=0.35\textwidth]{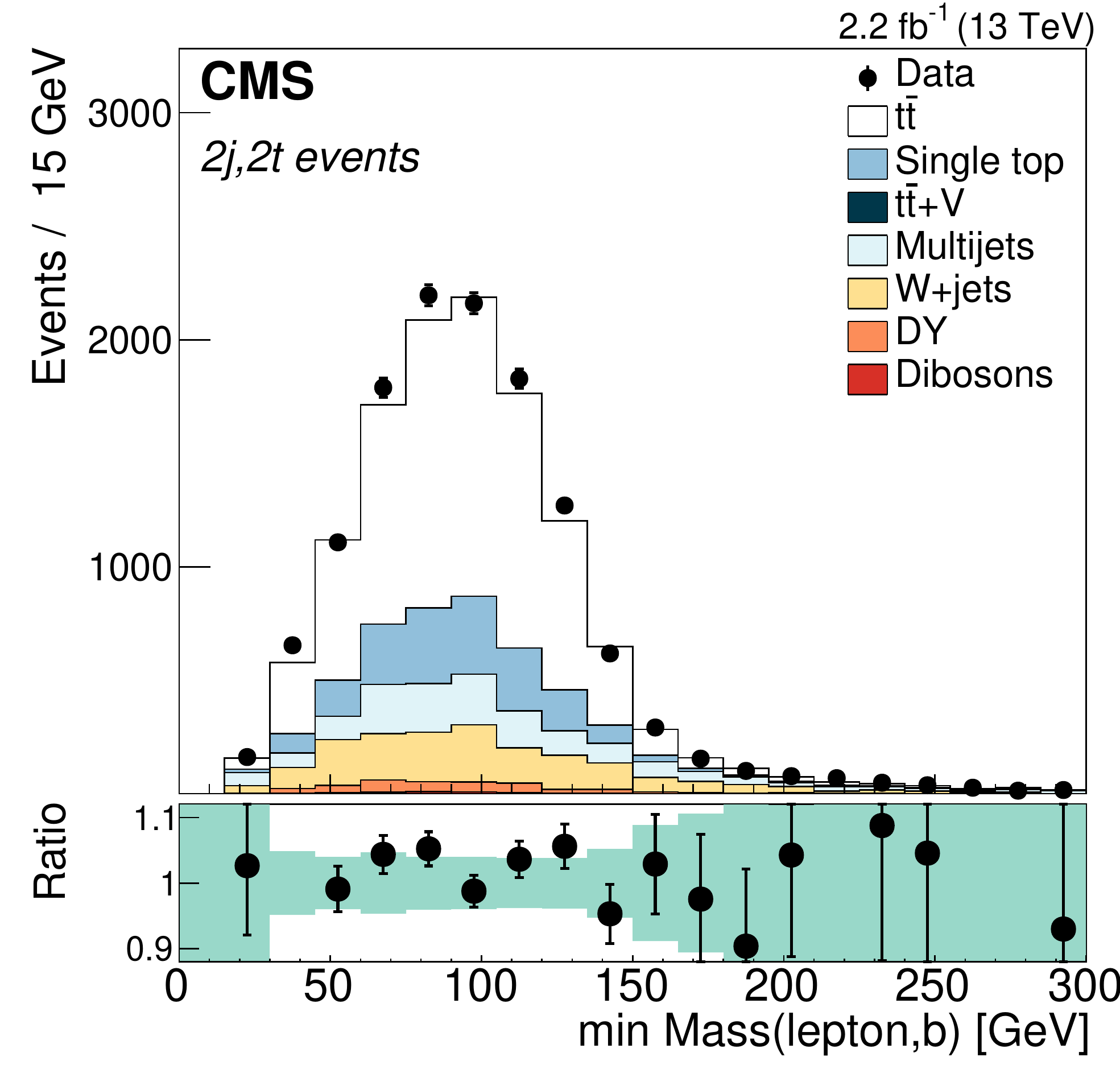}\\
\includegraphics[width=0.35\textwidth]{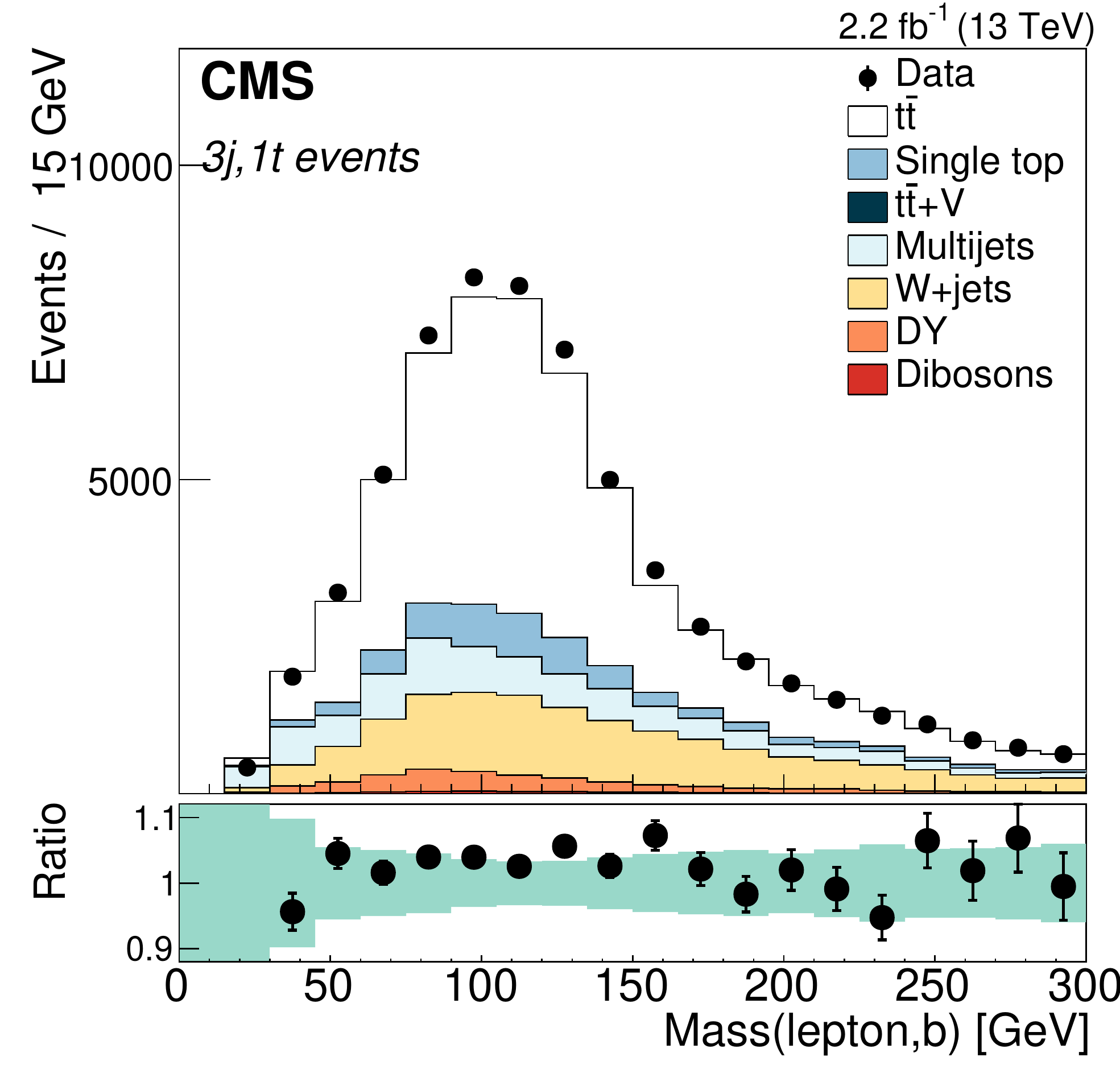}
\includegraphics[width=0.35\textwidth]{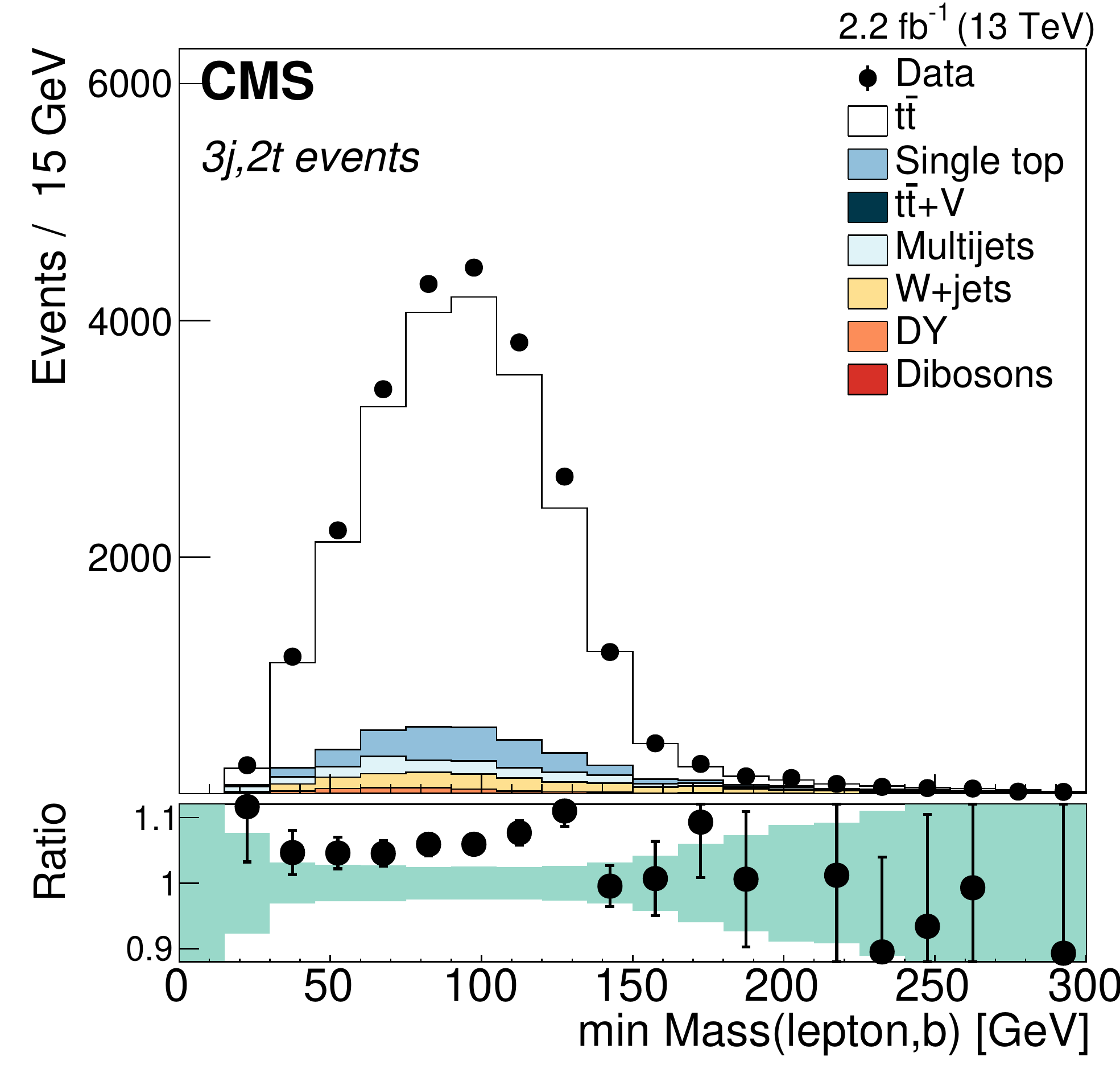}\\
\includegraphics[width=0.35\textwidth]{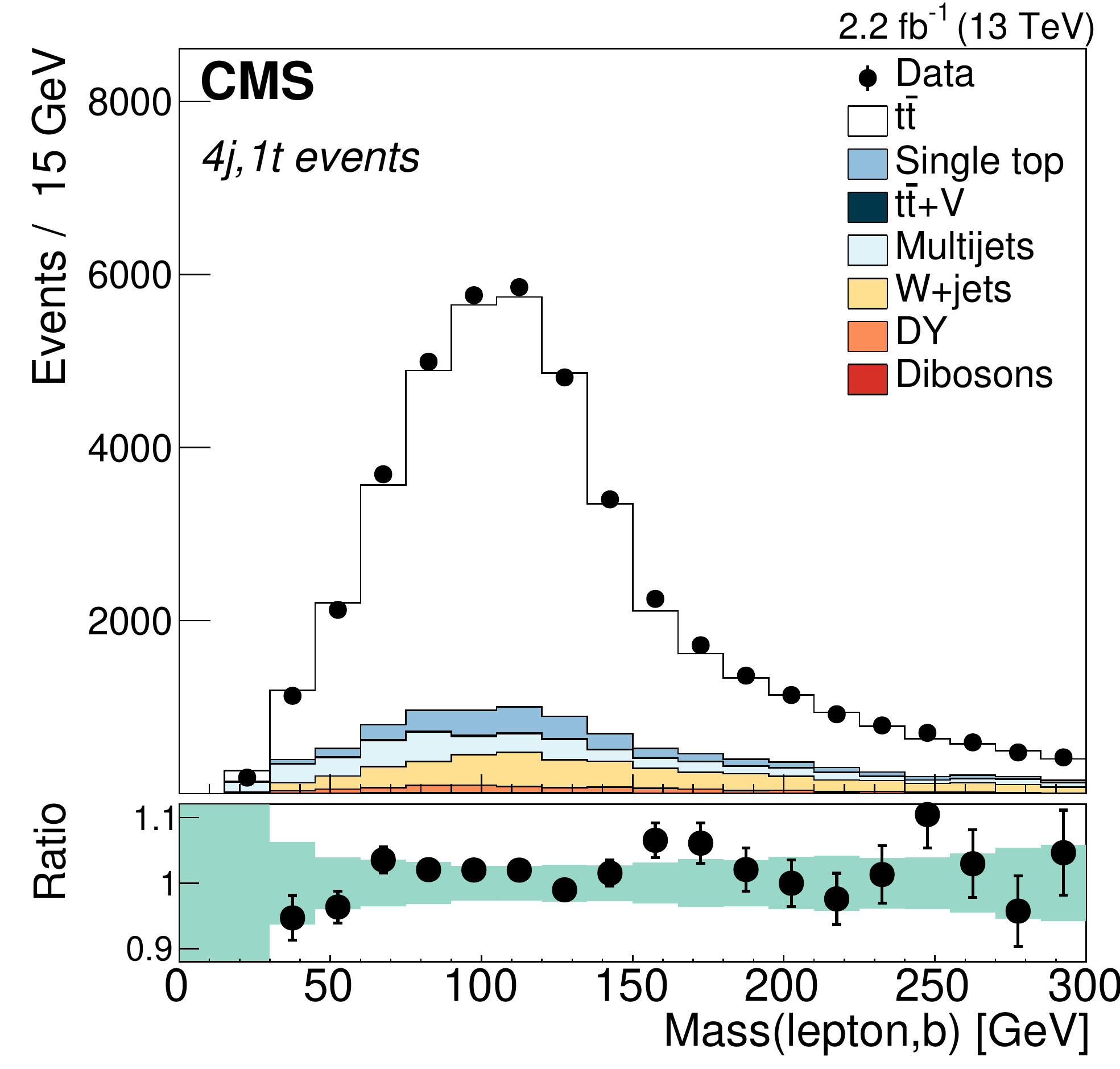}
\includegraphics[width=0.35\textwidth]{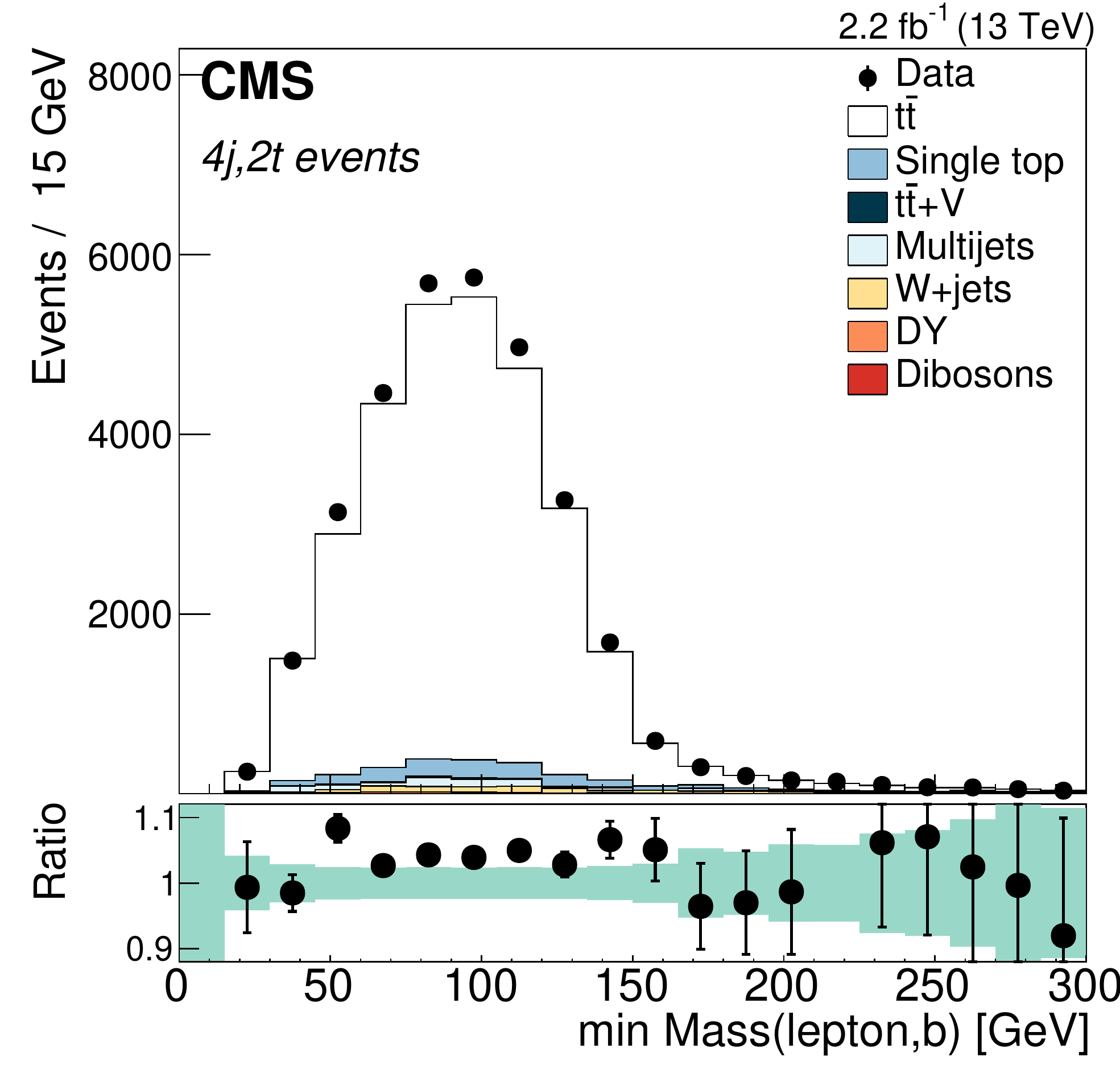}
\caption{
Distributions in the observables used to fit the data with the
contributions from all leptons and charges combined.
Panels on the left show the distributions in $M(\ell,\cPqb)$, and on the
right in $\min M(\ell,\cPqb)$, for events with one and two \cPqb-tagged
jets, respectively.
From top to bottom, the events correspond to those with 1, 2, 3, or at least 4 jets.
The lower plot in each panel shows the ratio between the data and expectations.
The relative uncertainty owing to the
statistical uncertainty in the simulations,
to the uncertainty in the normalization of the
contribution from multijet events
and to the systematic uncertainty in the total integrated
luminosity is represented as a shaded band.
}
\label{fig:obs}
\end{figure}

Figures~\ref{fig:obs} shows the $M(\ell,\cPqb)$ and $\min M(\ell,\cPqb)$ distributions for data, and the
expected contributions from signal and backgrounds in the various event categories.
When normalized
by the reference cross sections described in Section~\ref{subsec:datasim}
there is an overall good agreement between data and expectations.
The most noticeable differences
are related to the initial multijet background normalization
and the uncertainty in the \PW+jets normalization
which is improved
by the fitting procedure (see Section~\ref{sec:results}).

In the signal region, the agreement is good for the
simulation using the reference value $m_{\PQt}=172.5\GeV$.

The expectations for the rates and distributions
considered in the analysis are affected by different sources of
systematic uncertainties.
For each source, an induced variation can be parametrized,
and treated as a nuisance parameter in the fit
that is described in the next section.

Experimental uncertainties pertain mostly to the calibration of the
detector and to our assessment of its performance in the simulation.
The uncertainty in the efficiency of the trigger and the offline selection
is estimated by applying
different scale factors as a function of the \pt and $\eta$ of the isolated lepton.
The scale factors and their uncertainties are obtained using $\cPZ\to\ell\ell$ data,
based on a tag-and-probe method~\cite{Khachatryan:2010xn}.
The one standard deviation changes applied to the parameters of the simulated events
are typically on the order of 1--3\%.

The energy scales of the objects used in the analysis (leptons and
jets) are varied according to their estimated uncertainties. This can lead to a
migration of events to different categories because of the thresholds applied in the
preselection and the categorization of the events, as well as to
changes in the expected distributions of the observables.
When the energy scale of the leptons or jets
changes, it affects other variables (e.g., the missing momentum), which are
recomputed to reflect the new scales.
The uncertainty in the jet energy scale is subdivided into
independent sources. A total of 29 nuisance parameters related to the jet energy scale are included in the
fit described in the next section.
The parameters refer to the effect of uncertainties related to pileup, relative
($\eta$-dependent) calibration, high- and low-\pt extrapolation,
absolute-scale determination, and flavor-specific differences, amongst
others. The categories used for the jet energy scale
are similar to those used in the
$\sqrt{s}$ = 8\TeV analyses~\cite{Khachatryan:2016kdb,CMS-PAS-JME-15-001}.

The jet energy resolution is also affected by an uncertainty that is
estimated in our analysis by changing the simulated
resolution by one standard deviation as a function of the $\eta$ of the jet.
The corrections applied to the simulated \cPqb{}~jet, \cPqc{}~jet,
and light-flavor jet tagging efficiencies of the CSV algorithm
are changed according to their uncertainties~\cite{Chatrchyan:2012jua}. This also causes a
migration of events across the different \cPqb{} tagging categories
within the same jet multiplicity.
The uncertainty from the model used for the average pileup in the simulation
is estimated by implementing a
5\% change to the assumed inelastic pp cross section~\cite{Aaboud:2016mmw}.
Finally, a 2.3\% uncertainty is assigned to the estimated integrated
luminosity~\cite{CMS-PAS-LUM-15-001}.

For the estimate of the contribution from QCD multijet events we determine an
uncertainty owing to the normalization method of the nonisolated-lepton
sideband in data through an alternative scale factor obtained from
events with $M_\mathrm{T}<50\GeV$, where $M_\mathrm{T}$ is the transverse
mass computed from the lepton candidate and the missing momentum of
the event.
This yields an intrinsic uncertainty of
$\approx$30--60\%,
depending on the category. Furthermore,
uncertainties in the distributions of events caused by
the normalizations of other than multijet contributions are
obtained by changing the individual sources in the control
regions by $\pm$30\%.
These uncertainties are considered
uncorrelated across all categories of the analysis.

Theoretical uncertainties affect the predictions for the acceptance and
the distributions in the signal and nonmultijet background processes.
We consider independent changes in $\mu_\mathrm{R}$ or $\mu_\mathrm{F}$ in the
\ttbar, \PW+jets, and t\PW{} processes by factors of 2 and 1/2.
For the signal, we estimate the parton shower uncertainty
by using alternative \POWHEG+\PYTHIA 8 samples, with
the parton shower scale value changed by factors of 2 and 1/2.
This affects the fragmentation and hadronization of
the jets initiated by the matrix-element calculation, as well as the
emission of extra jets.
The variation in the acceptance and distributions
obtained by using \HERWIGpp instead of \PYTHIA~8
to interface the \POWHEG generator is included as a systematic
uncertainty in the modeling of \ttbar in the fit.
An additional uncertainty is assigned based on the difference found
between the \POWHEG and \textsc{MG5}\_a\MCATNLO simulations.

For the signal, we also consider an uncertainty
in the \pt distribution of the top quark, based on the CMS measurements at
$\sqrt{s}=8$~\cite{Khachatryan:2015oqa}
and 13\TeV~\cite{Chatrchyan:2013boa}.
The simulation is reweighted using a data-to-simulation scale
factor that is verified to be consistent with the measurements
performed in both data sets, and the difference is used to assign the
uncertainty in the modeling of the top quark \pt.

Uncertainties in the modeling of the single top quark background include
changes
of $\mu_\mathrm{R}/\mu_\mathrm{F}$ for the \PQt{} and \PQt\PW{} channels.
At NLO QCD,  \PQt\PW{} production is expected to interfere with \ttbar
production, owing to the similar initial and final states of some
diagrams~\cite{Frixione:2008yi,Belyaev:1998dn,White:2009yt}.
Two schemes for defining the \PQt\PW{} signal that distinguish it from
\ttbar production have therefore been compared in this analysis:
 the ``diagram removal''  method~\cite{Frixione:2008yi}, in which all doubly-resonant NLO \PQt\PW{} diagrams are removed,
and the  ``diagram subtraction''  scheme~\cite{Frixione:2008yi, Tait:1999cf}, where a gauge-invariant subtraction
term modifies the NLO \PQt\PW{} cross section to locally cancel the contribution from \ttbar.
In addition to the theoretical uncertainties described above, all background processes are assigned their corresponding
theoretical  uncertainties in their normalization.

\section{Fitting procedure and results}
\label{sec:results}

The \ttbar production cross section is measured by performing a
maximum-likelihood fit to
the number of events counted in the different categories.
The likelihood function takes into account the expectations
for contributions from
different background processes as well as signal.
The expectations for signal and backgrounds
depend on:
(i) the simulation- or data-based expectations ($\hat{\mathrm{S}}$ or
$\hat{\mathrm{B}}$ for
signal and background, respectively),
and (ii) nuisance parameters  ($\theta_i$) that
reflect the uninteresting variables used to
control the effect
of the systematic variations
described in the previous section.
The effect of each source of uncertainty is separated in a rate-changing
and shape-changing nuisance parameter.
In the fit, the nuisance parameters are assumed to be distributed according to
log-normal probability distribution functions (pdfs) if
affecting the rate, or Gaussian pdfs if
affecting the shapes.
We denote generally the pdfs associated with a nuisance parameter
as $\rho(\theta_i)$.
The signal expectation is
also modulated by a multiplicative factor,
which is defined by the ratio of the measured cross section to the reference theoretical value,
i.e., the signal strength $\mu=\sigma/\sigma_\text{th}$ for $m_\PQt=172.5\GeV$.
For each category (k), we write the total number of expected events as:
\begin{equation}
\hat{N}_k(\mu,\Theta)
= \mu \, \hat{\mathrm{S}}_k  \, \prod_i (1+\delta_i^\mathrm{S}\theta_i)+
\hat{\mathrm{B}}_k  \, \prod_i
(1+\delta_i^\mathrm{B}\theta_i)~,
\label{eq:nexp}
\end{equation}
where $\Theta$ is the set of all nuisance parameters,
the index k runs over the bins of the distributions
(or the counts in different event categories for the cross-check analysis),
and $\delta_i^\mathrm{S}$ and
$\delta_i^\mathrm{B}$ are changes in
yields induced
through one-standard-deviation changes in
the {\it i}$^\text{th}$ sources of uncertainty in the signal
and backgrounds, respectively.
The likelihood function is defined as:
\begin{equation}
\mathcal{L}(\mu,\Theta) =
\prod_k \mathcal{P} \left[ N_k | \hat{N}_k(\mu,\theta_i) \right]\,
\prod_i \rho(\theta_i),
\label{eq:ll}
\end{equation}
where $\mathcal{P}$ is a Poisson distribution and $N_k$ is the number of events
observed in the $k$th category.
The cross section is measured by maximizing the profile likelihood ratio
(PLR) test statistic:
\begin{equation}
\lambda(\mu) = \frac{\mathcal{L}(\mu,\hat{\hat\Theta})}{\mathcal{L}(\hat{\mu},\hat\Theta)},
\label{eq:profll}
\end{equation}
where the quantities $\hat{\hat\Theta}$ correspond to the set of nuisance parameter values
$\theta_i$
that maximize the likelihood for the specified signal strength (also
known as the conditional likelihood),
and $\hat\mu$, $\hat\Theta$ are respectively the values of $\mu$ and
the set of $\theta_i$
that maximize the likelihood.
In the presence of nuisance parameters, the resulting PLR as a function of $\mu$ tends to be broader
relative to the one obtained when the values are well known and fixed. This reflects the loss of information in $\mu$
because of the presence of
systematic uncertainties~\cite{Cowan:2010js}.

Although $m_\PQt$
does not contribute an intrinsic uncertainty
in the measurement of the cross section, 
since the $M(\ell,\cPqb)$ distribution is used in the fit, its shape
has a direct dependence on $m_\PQt$ that needs to be taken into account.
We thus include in the fit a parameterization of the effect of varying $m_\PQt$ by ${\pm}3\GeV$
while measuring the cross section as the parameter of interest.
This parameterization is performed for 
both the signal and the single top quark simulations.
With this procedure, the fit accomodates for a possibly different value of $m_\PQt$
than that assumed by default in the simulation
but witout correlating this with the pole mass to be extracted from the inclusive \ttbar production rate,
as originally proposed in~Ref.~\cite{Kieseler:2015jzh}.

\begin{figure}[htbp]
\includegraphics[width=0.49\textwidth]{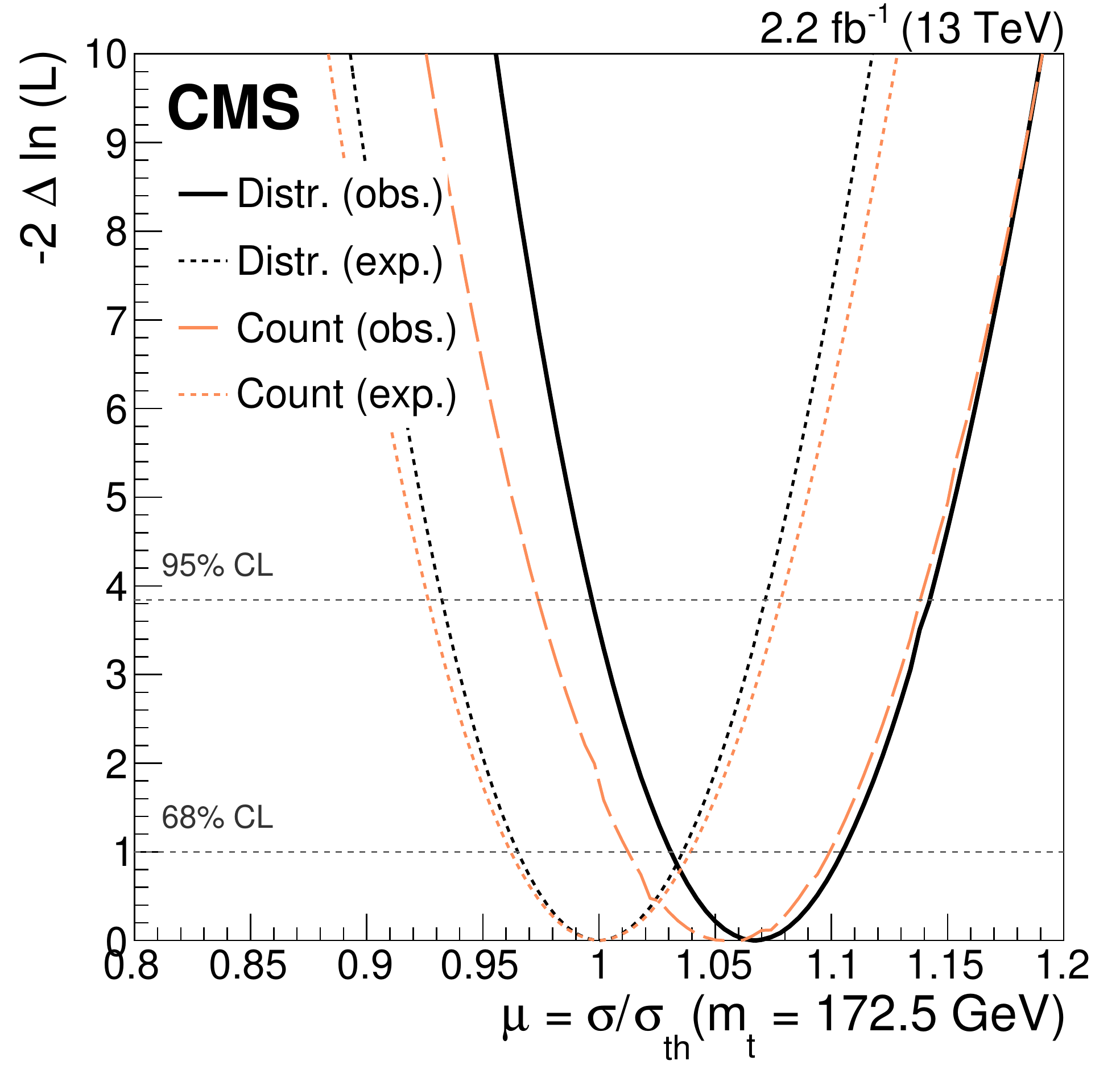}
\includegraphics[width=0.49\textwidth]{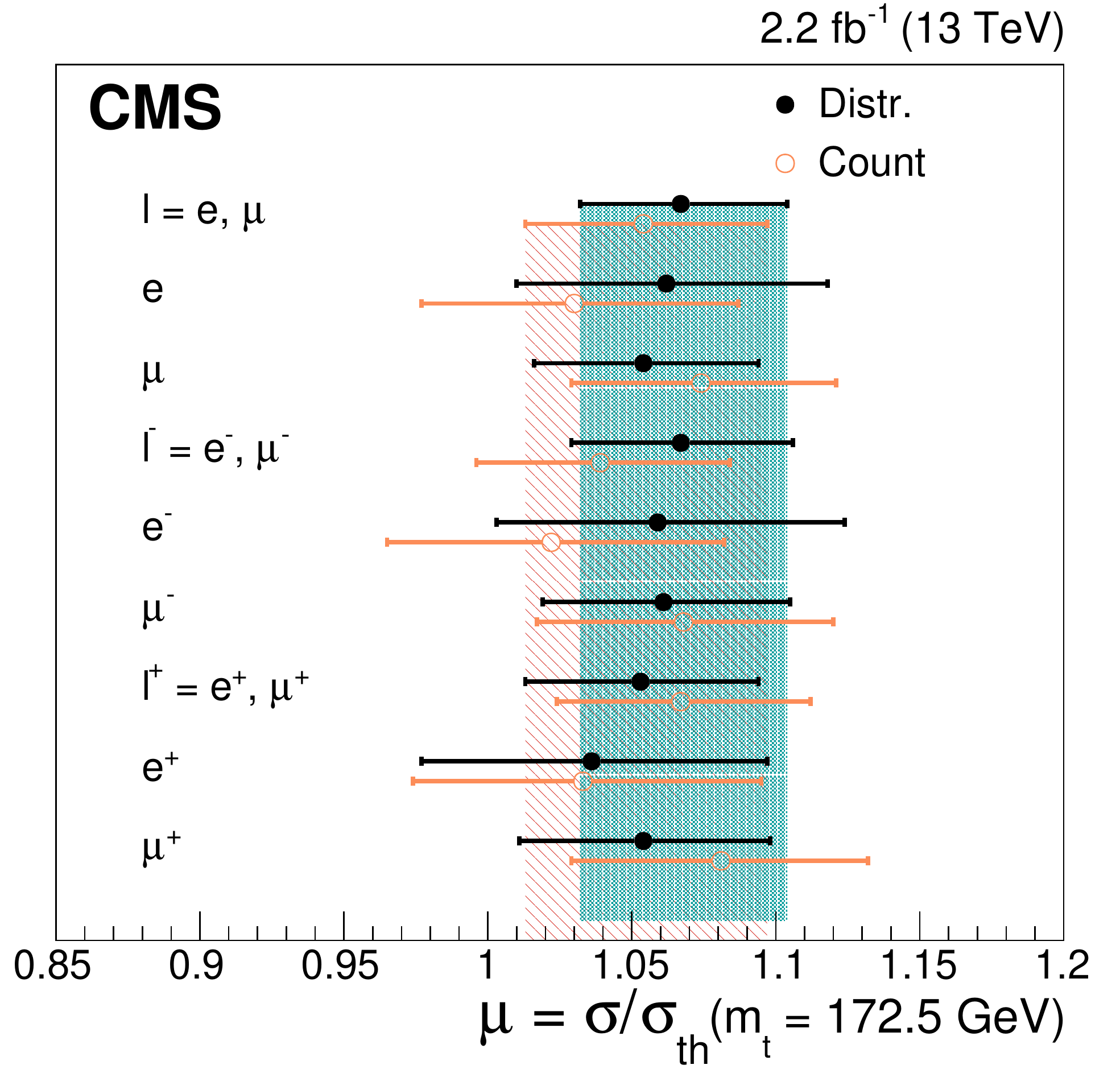}
\caption{(left) The observed (solid curve) and expected (dashed curve)
  variation
of the likelihood as a function of the signal strength
$\mu$ for the distribution-based analysis.
The expected curve is obtained by performing the fit using simulated
events with $m_\PQt=172.5\GeV$.
For comparison, the corresponding curves for the counting
cross-check analysis are also shown.
The two horizontal lines represent the values
in the PLR that are used to determine the 68\% and 95\% confidence level
(CL) intervals for the signal strength.
(right) Comparison of the values of the signal strength extracted for
different combinations of events for the distribution-based default
analysis
(solid circles) and the cross-check counting analysis (open circles).
The horizontal bars represent the
total uncertainties, except the beam energy uncertainty.
The shaded bands represent the uncertainty in the final combined
signal strength obtained from the distribution-based and cross-check analyses.}
\label{fig:results}
\end{figure}

Figure~\ref{fig:results}\,(left) shows the variation of the likelihood as a function
of the signal strength from the data and the expected variation from the simulation.
From the fit, we measure $\mu=1.067 \pm 0.002\stat~^{+0.037}_{-0.035}\syst$.
The \ttbar cross section in the visible phase space is thus measured
with a total uncertainty of 3.4\%.
As a check, the Monte Carlo simulated signal and background events corresponding to the same
integrated luminosity as the data are used as pseudo-data with
$m_\PQt=172.5\GeV$ in the fit. The resulting value of the signal
strength is $\mu=1.000 \pm 0.002\stat~^{+0.035}_{-0.034}\syst$.
This is the expected value of $\mu$, and the agreement of the statistical
and systematic uncertainties with those from the fit to the data
is a good check on the fitting procedure.

The default analysis using the shapes of the distributions (labeled ``Distr.'')
is also compared with a simpler cross-check analysis (labeled
``Count").
The cross-check analysis does not use kinematic information,
but uses the number of events in the different jet and \cPqb-tagged jet categories, and
the expected yields. The two results are in agreement with each
other, with the cross-check analysis having a larger uncertainty:
$\mu=1.054\pm 0.002\stat~^{+0.043}_{-0.041}\syst$.

The post-fit normalizations for the main backgrounds
(\PW+jets and multijets) tend to be higher by 1--6\% in the main analysis
with respect to those from the cross-check analysis. This results in a different
signal strength between the two analyses.

Figure~\ref{fig:results}\,(right) compares the inclusive $\mu$ result
for both the default and cross-check analyses (top set of points)
with the corresponding values for the different lepton charges and flavors.
The results are found to be consistent with each other in the different combinations.

\begin{figure}[htb]
\centering
\includegraphics[width=0.7\textwidth]{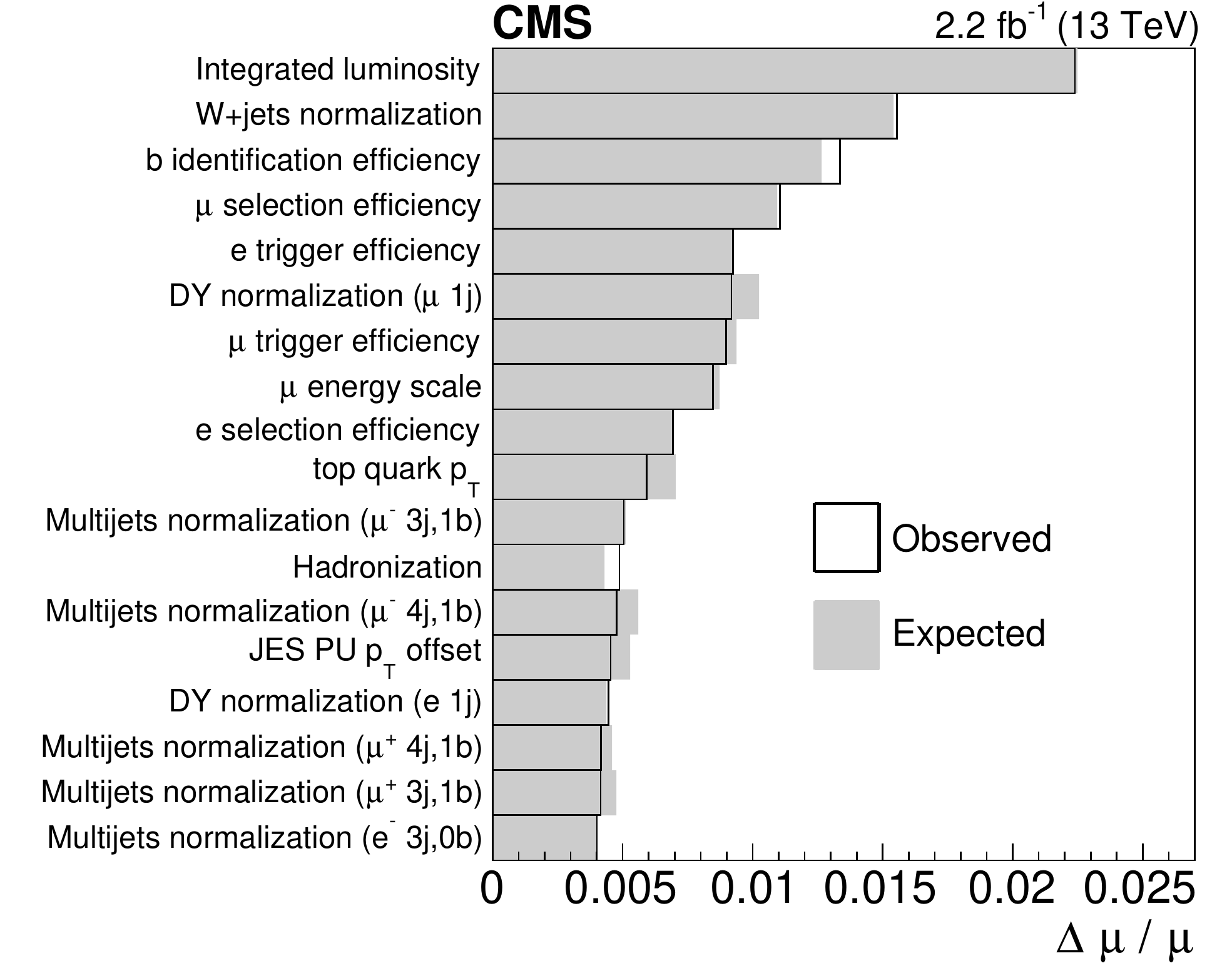}
\caption{
Estimated change $\Delta\mu$ in the measured signal strength $\mu$, coming
from the listed experimental and theoretical sources of uncertainties
in the main analysis. The open bars represent the values of the
observed impact relative to the fitted signal strength.
The values are  compared to the expectations (shaded bars) by performing the fit using simulated
events with $m_\PQt=172.5\GeV$.
The various contributions are shown from the largest to the smallest observed impact.
}
\label{fig:impacts}
\end{figure}

The impact of the sources of uncertainty in the fit is evaluated by
making use of the set of post-fit values of the nuisance parameters,
and computing the shift induced in the signal strength as each nuisance parameter is
fixed at its $\pm 1$ standard deviation post-fit value,
with all other parameters profiled as normal.
By repeating the fits, the effect of some nuisance parameters being fixed may be reabsorbed
by a variation of the ones being profiled, owing to correlations.
Figure~\ref{fig:impacts} summarizes the values obtained for the leading sources of uncertainty in the fit.
The dominant sources of uncertainty in
both analyses are related to the integrated luminosity, trigger and selection efficiencies,
and the model of the \PW{}+jets background.
These are expected to impact the signal strength at the level of 1--2.5\%
The analysis of the distributions is effectively able
to mitigate most uncertainties related to the modeling of \ttbar.
The modeling of the top quark \pt
and the choice of the hadronizer are the dominant signal modeling uncertainties
but their impact in the fit is observed to be $<$1\%.
Uncertainties related to the modeling of the multijets background
are observed to impact the fit at the level of $<$0.5\%
None of the nuisance parameters used in the fit is observed to be significantly
pulled from its initial value and its behavior is similar to that expected
by performing the fit using simulated
events with $m_\PQt=172.5\GeV$.
Nuisance parameters related to the integrated luminosity and the trigger and selection efficiencies
are observed not to be constrained in the fit procedure.

The signal strength is measured in a region of phase space where
the lepton has $\pt>30\GeV$ and $\abs{\eta}<2.1$, and at least one jet
has $\pt>30\GeV$ and $\abs{\eta}<2.5$.
The resulting visible \ttbar cross section in this phase-space region is determined
to be
\begin{equation*}
\sigma^\text{vis}_{\ttbar}= 208.2 \pm 0.4 \stat~_{-4.9}^{+5.5}\syst \pm 4.8\lum\unit{pb},
\end{equation*}
where the last uncertainty is from the integrated luminosity.

{\tolerance=1400
The extrapolation to the full phase space is performed by using the acceptance
estimated from the \ttbar simulation.
Using \POWHEG, we determine the acceptance to be $0.2345 \pm 0.0001\stat~^{+0.0044}_{-0.0043}\syst$,
where the systematic uncertainty comes
from changing $\mu_\mathrm{R}/\mu_\mathrm{F}$ ($\pm0.0017$),
considering the CT14 PDF and $\alpha_\mathrm{S}$ uncertainties ($^{+0.0009}_{-0.0007}$)~\cite{Dulat:2015mca},
and changing the parton shower algorithm used to interface with the matrix-element generator,
\ie, \PYTHIA8 vs. \HERWIGpp, ($\pm0.0039$).
The total uncertainty associated with the extrapolation is
estimated to be 1.6\%. This uncertainty is added in quadrature to the
systematic uncertainty obtained in the fitted fiducial region when
extrapolating the measurement to the full phase space.
\par}

Summing the statistical (0.2\%),
systematic (3.0\%), and
integrated luminosity (2.3\%) uncertainties in quadrature,
we obtain a total relative uncertainty in the \ttbar cross section of
3.9\%. The final result is:
\begin{equation*}
\sigma_{\ttbar}= 888 \pm 2 \stat ~_{-26}^{+28}\syst \pm 20\lum\unit{pb},
\end{equation*}
in agreement with the NNLO+NNLL prediction~\cite{top++} and the measurement
derived from analyzing events in the electron + muon final state from the
same data set~\cite{Khachatryan:2016kzg}.

The result can be reinterpreted to extract the pole mass $m_\PQt$ of the top quark
by using the dependence of the cross section on this
parameter.
We make use of the \textsc{top++} program ~\cite{top++} and the
CT14 NNLO PDF~\cite{Dulat:2015mca} to
parametrize the dependence of the cross section on the top quark mass.
The parametrization used is:
\begin{equation}
\sigma(m_{\PQt})=\sigma(m_\text{ref})\,\left(\frac{m_{\rm  ref}}{m_\PQt}\right)^4 \, \left[ 1+a_1 (\frac{m_\PQt}{m_\text{ref}}-1)+a_2(\frac{m_\PQt}{m_\text{ref}}-1)^2\right],
\label{eq:xsecpolemass}
\end{equation}
where $m_\text{ref}=172.5\GeV$ is the reference mass value, and $a_1$ and
$a_2$ are coefficients determined after
performing the calculations with
various $m_\PQt$ hypotheses.
The  effects induced by the choice of $\mu_\mathrm{R}/\mu_\mathrm{F}$,
the uncertainty in the PDF+$\alpha_\mathrm{S}$, and
uncertainties in the beam energy, are evaluated by
recomputing the cross section after changing these parameters
within their uncertainties. The resulting typical uncertainties
in $\sigma(m_t)$ amount to $^{+2.5\%}_{-3.7\%}$, $^{+2.7\%}_{-2.6\%}$, and $\pm$0.23\%,
respectively.
The latter reflects a $\pm$0.1\% uncertainty in the
beam energy at which the data have been collected~\cite{Wenninger:2254678}.

To measure the pole mass, the likelihood function (Eq.~(\ref{eq:ll})) is reparametrized,
transforming $\mu$ into a functional form that depends on the
top quark mass
\begin{equation}
\mu(m_\PQt)=\frac{\sigma(m_\PQt)}{\sigma_\text{th}} \, \frac{A}{A (m_\PQt)},
\label{eq:mumpole}
\end{equation}
where the last factor (${A}/{A (m_\PQt)}$), is a mass-dependent
correction to the acceptance.
Using simulated \ttbar samples with different $m_\PQt$, we
find that the acceptance changes by 0.08\% per $\Delta m_{\PQt}=1\GeV$.

The uncertainty in the extrapolation, as well as the theoretical uncertainties
that affect the parameterization as a function of  $m_\PQt$
coming from the choices of $\mu_\mathrm{R}/\mu_\mathrm{F}$,
PDF, $\alpha_\mathrm{S}$, and beam energy, are added as
extra nuisance parameters in the fit for the pole mass.
With the exception of $\mu_\mathrm{R}/\mu_\mathrm{F}$, which
is defined through a
log-uniform probability distribution consistent with the procedure adopted in~Ref.~\cite{Chatrchyan:2013haa},
the remaining uncertainties are assigned a log-normal function.
After repeating the maximum-likelihood fit, we obtain
\begin{equation*}
m_\PQt = 170.6\pm2.7\GeV,
\end{equation*}
where the quoted uncertainty  contains both statistical and systematic contributions.
The result agrees with that obtained
using the NNPDF3.0 NNLO PDF~\cite{Ball:2014uwa}:
$m_\PQt = 170.3~^{+2.6}_{-2.7}\GeV$. The latter is only used as a
cross-check as the
NNPDF3.0 PDF includes top-quark-related data in the determination of
the proton PDFs.
In both cases, the best-fit value is determined by fixing the nuisance parameter
associated
with the choice of the  $\mu_\mathrm{R}$ and $\mu_\mathrm{F}$ ratio
to its post-fit value, and repeating the scan
of the likelihood. This procedure is adopted to resolve the
almost degenerate behavior of the likelihood, induced
through the use of a log-uniform pdf assigned to the choice of
the $\mu_\mathrm{R}$ and $\mu_\mathrm{F}$ ratio.

\begin{figure}[!htb]
\centering
\includegraphics[width=0.65\textwidth]{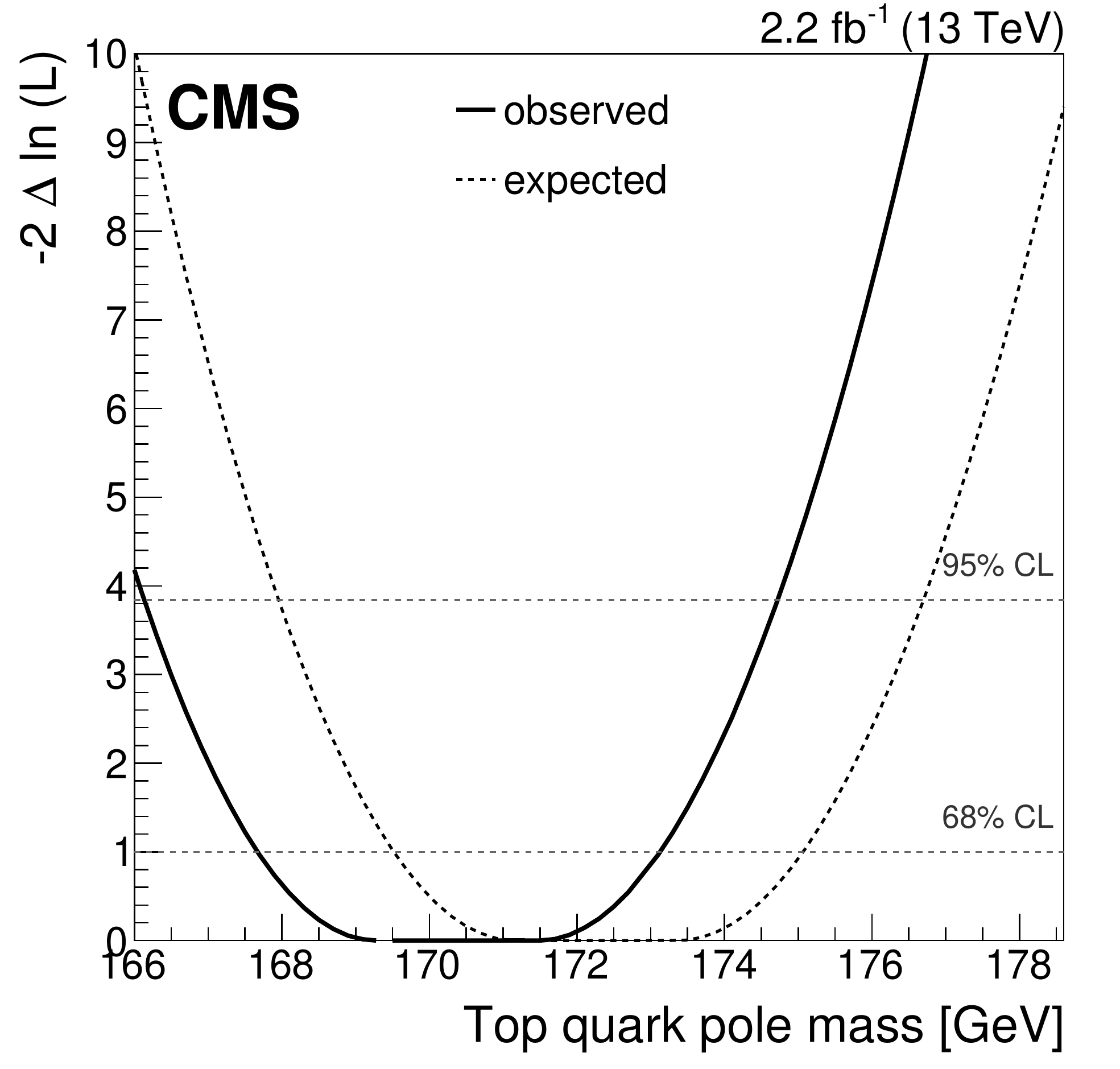}
\caption{Dependence of the likelihood on the top quark pole mass (solid curve).
The expected dependence from the simulation, using the a priori set of
nuisance parameters with their expected values at  $m_\PQt=172.5\GeV$,
is shown for comparison as the dotted curve.
The changes in the likelihood corresponding to the 68\% and 95\% confidence levels (CL)
are shown by the dashed lines.}
\label{fig:mpolell}
\end{figure}

Figure~\ref{fig:mpolell} shows the variation of the likelihood as a function
of the top quark pole mass. For comparison, the expected likelihood
from the Asimov set of nuisance parameters at $m_\PQt=172.5\GeV$ is
shown.

The impact of each source of systematic uncertainty in the values corresponding to the fit is estimated using a
similar procedure to the one described above for the cross section measurement.
Table~\ref{tab:mpoleimpacts} summarizes the estimated uncertainties in the determination of $m_\PQt$ from the
measured cross section.

\begin{table}[!htb]
\centering
\topcaption{The source and value of the systematic uncertainties in
  the measurement of $m_\PQt$. \label{tab:mpoleimpacts}}
\begin{tabular}{lll}
Source & $\Delta m_\PQt\,[\GeVns{}]$ \\
\hline
Uncertainties from the fit in the fiducial region & $-2.2$ /$+2.5$ \\
Extrapolation to the full phase space & $-0.7$ /$+1.1$ \\
Beam energy & $-0.08$ /$+0.12$ \\
$\mu_\mathrm{R}/\mu_\mathrm{F}$ and PDF+$\alpha_\mathrm{S}$ & $-0.9$ /$+1.1$ \\
\hline
Total & $\pm$2.7 \\
\end{tabular}
\end{table}

\section{Summary}
\label{sec:summary}

A measurement of the  $\ttbar$ production cross section at
  $\sqrt{s} = 13$\TeV has been presented by CMS in final states containing one isolated lepton and
  at least one jet. The acceptance in the fiducial part of the phase space is estimated
  with an uncertainty of 1.6\% and has a negligible dependence on
  $m_\PQt$. By performing a simultaneous fit to event
  distributions in 44 independent categories, we measure the strength of the \ttbar
  signal relative to the NNLO+NNLL~\cite{top++} computation with an uncertainty
  of 3.9\%. We obtain an inclusive \ttbar production
  cross section
  $\sigma_{\ttbar}= 888 \pm 2 \stat ~_{-26}^{+28}\syst \pm 20\lum\unit{pb}$,
  which is compatible with the standard model prediction,
  competing in precision with it~\cite{top++} and with similar measurements of this quantity
  at the same $\sqrt{s}$~\cite{Khachatryan:2015uqb,Aaboud:2016pbd,Khachatryan:2016kzg}.
  In addition, the top quark pole mass, $m_\PQt$, is extracted at NNLO using the same data and the CT14 PDF set and found to be
  $m_\PQt=170.6\pm2.7\GeV$. This value is in good agreement with measurements using other techniques.

\begin{acknowledgments}
\label{sec:ack}

We congratulate our colleagues in the CERN accelerator departments for
the excellent performance of the LHC and thank the technical and
administrative staffs at CERN and at other CMS institutes for their
contributions to the success of the CMS effort. In addition, we
gratefully acknowledge the computing centres and personnel of the
Worldwide LHC Computing Grid for delivering so effectively the
computing infrastructure essential to our analyses. Finally, we
acknowledge the enduring support for the construction and operation of
the LHC and the CMS detector provided by the following funding
agencies: BMWFW and FWF (Austria); FNRS and FWO (Belgium); CNPq,
CAPES, FAPERJ, and FAPESP (Brazil); MES (Bulgaria); CERN; CAS, MoST,
and NSFC (China); COLCIENCIAS (Colombia); MSES and CSF (Croatia); RPF
(Cyprus); SENESCYT (Ecuador); MoER, ERC IUT, and ERDF (Estonia);
Academy of Finland, MEC, and HIP (Finland); CEA and CNRS/IN2P3
(France); BMBF, DFG, and HGF (Germany); GSRT (Greece); OTKA and NIH
(Hungary); DAE and DST (India); IPM (Iran); SFI (Ireland); INFN
(Italy); MSIP and NRF (Republic of Korea); LAS (Lithuania); MOE and UM
(Malaysia); BUAP, CINVESTAV, CONACYT, LNS, SEP, and UASLP-FAI
(Mexico); MBIE (New Zealand); PAEC (Pakistan); MSHE and NSC (Poland);
FCT (Portugal); JINR (Dubna); MON, RosAtom, RAS, RFBR and RAEP
(Russia); MESTD (Serbia); SEIDI and CPAN (Spain); Swiss Funding
Agencies (Switzerland); MST (Taipei); ThEPCenter, IPST, STAR, and
NSTDA (Thailand); TUBITAK and TAEK (Turkey); NASU and SFFR (Ukraine);
STFC (United Kingdom); DOE and NSF (USA).

Individuals have received support from the
Marie-Curie programme and the European Research Council and EPLANET
(European Union); the Leventis Foundation; the A. P. Sloan Foundation;
the Alexander von Humboldt Foundation; the Belgian Federal Science
Policy Office; the Fonds pour la Formation \`a la Recherche dans
l'Industrie et dans l'Agriculture (FRIA-Belgium); the Agentschap voor
Innovatie door Wetenschap en Technologie (IWT-Belgium); the Ministry
of Education, Youth and Sports (MEYS) of the Czech Republic; the
Council of Science and Industrial Research, India; the HOMING PLUS
programme of the Foundation for Polish Science, cofinanced from
European Union, Regional Development Fund, the Mobility Plus programme
of the Ministry of Science and Higher Education, the National Science
Center (Poland), contracts Harmonia 2014/14/M/ST2/00428, Opus
2014/13/B/ST2/02543, 2014/15/B/ST2/03998, and 2015/19/B/ST2/02861,
Sonata-bis 2012/07/E/ST2/01406; the Thalis and Aristeia programmes
cofinanced by EU-ESF and the Greek NSRF; the National Priorities
Research Program by Qatar National Research Fund; the Programa
Clar\'in-COFUND del Principado de Asturias; the Rachadapisek Sompot
Fund for Postdoctoral Fellowship, Chulalongkorn University and the
Chulalongkorn Academic into Its 2nd Century Project Advancement
Project (Thailand); and the Welch Foundation, contract C-1845.
\end{acknowledgments}
\bibliography{auto_generated}
\cleardoublepage \appendix\section{The CMS Collaboration \label{app:collab}}\begin{sloppypar}\hyphenpenalty=5000\widowpenalty=500\clubpenalty=5000\textbf{Yerevan Physics Institute,  Yerevan,  Armenia}\\*[0pt]
A.M.~Sirunyan, A.~Tumasyan
\vskip\cmsinstskip
\textbf{Institut f\"{u}r Hochenergiephysik,  Wien,  Austria}\\*[0pt]
W.~Adam, E.~Asilar, T.~Bergauer, J.~Brandstetter, E.~Brondolin, M.~Dragicevic, J.~Er\"{o}, M.~Flechl, M.~Friedl, R.~Fr\"{u}hwirth\cmsAuthorMark{1}, V.M.~Ghete, C.~Hartl, N.~H\"{o}rmann, J.~Hrubec, M.~Jeitler\cmsAuthorMark{1}, A.~K\"{o}nig, I.~Kr\"{a}tschmer, D.~Liko, T.~Matsushita, I.~Mikulec, D.~Rabady, N.~Rad, B.~Rahbaran, H.~Rohringer, J.~Schieck\cmsAuthorMark{1}, J.~Strauss, W.~Waltenberger, C.-E.~Wulz\cmsAuthorMark{1}
% ...
\vskip\cmsinstskip
% ...
1:~~Also at Vienna University of Technology, Vienna, Austria\\
% ...

\end{sloppypar}
\end{document}